# Group Representations, Error Bases and Quantum Codes

E. Knill

knill@lanl.gov, Mail Stop B265
Los Alamos National Laboratory
Los Alamos, NM 87545.

## Abstract

This report continues the discussion of unitary error bases and quantum codes begun in [8]. Nice error bases are characterized in terms of the existence of certain characters in a group. A general construction for error bases which are non-abelian over the center is given. The method for obtaining codes due to Calderbank et al. [2] is generalized and expressed purely in representation theoretic terms. The significance of the inertia subgroup both for constructing codes and obtaining the set of transversally implementable operations is demonstrated.
**Note:** This report is preliminary. Please contact the author if you wish to be notified of updates.

## 1 Overview

This report discusses the construction of quantum codes based on nice error bases [8]. The main conclusion is that much of the relevant theory can be cast in terms of representations of finite groups. It is shown that nice error bases are equivalent to the existence of an irreducible character with non-zero values only on the center. The technique for obtaining codes based on "eigenspaces" of representations of abelian groups due to Calderbank et al. [2] is cast in terms of projection operators derived from the characters. This yields two potential generalizations of the construction to non-abelian subgroups, one exploiting the characters, the other exploiting primitive idempotents. The two constructions yield different sets of correctable or detectable errors. In both cases, the operators of the error basis which do not belong to the detectable set are in the inertia subgroup of a representation. The question of how to encode and decode codes using these constructions is not addressed. The inertia subgroup is also critical in constructing the syndrome space and in determining which operations can easily be implemented transversally (i.e. fault tolerantly). A simple way of exploiting classical codes over the finite field with $p^d$ elements is given. This method uses a non-canonical error basis on $d$ systems with $p$ states each.

This manuscript should be viewed as a continuation of [8]. It is assumed that the reader is familiar with the terms introduced there. For the necessary background on the representation theory of finite groups, see any advanced text on group theory such as [3, 6].

## 2 Characterization of Nice Unitary Error Bases

A nice unitary error basis on a Hilbert space $\mathcal{H}$ of $n$ dimensions is defined as a set $\mathcal{E} = \{E_g\}_{g \in G}$ where $E_g$ is unitary on $\mathcal{H}$, $G$ is a group of order $n^2$, $\text{tr} E_g = n\delta_{g,1}$ and $E_g E_h = \omega_{g,h} E_{g*h}$. By renormalizing the operators of the error basis, it can be assumed that $\det E_g = 1$, in which case $\omega_{g,h}$ is an $n$'th root of unity. Error bases with this property are called *very nice*. Such error bases generate a finite group of unitary operators $\bar{\mathcal{E}}$ whose center consists of scalar multiples of the identity. An *error group* is a finite group of unitary operators generated by a nice unitary error basis and multiples of the identity. The group $H$ is an *abstract error group* if it is isomorphic to an error group.

**Theorem 2.1.** *The finite group $H$ is an abstract error group iff $H$ has an irreducible character supported on the center and the kernel of the associated irreducible representation is trivial.*

*Proof.* Let $H$ be isomorphic to an error group of a nice error basis $\mathcal{E}$. The isomorphism induces an irreducible representation of $H$. The trace condition on the error group implies that the irreducible character $\chi$ of the representation is non-zero only at multiples of the identity. This is the center of the error group. The kernel of the representation is trivial by construction.

Suppose that $H$ has an irreducible character $\chi$ which is supported in the center $Z$ of $H$. Let $\sigma$ be the associated unitary representation with trivial kernel. Since the representation is irreducible, $\sigma(Z)$ consists of multiples of the identity. Let $G = H/Z$. For each $g \in H/Z$, pick a representative $h(g)$ in $H$, choosing $h(1) = 1$. Let $E_g = \sigma(h(z))$. Then $E_g E_h = \sigma(c) E_{g*h}$ for some $c \in Z$. We have $\sigma(c) = \omega_c I$. Since the character is zero except on the center, $\text{tr} E_g \propto \delta_{g,1}$. Thus $\text{tr}(E_g^\dagger E_h) = \xi(E_g^{-1} E_h) \propto \delta_{g,h}$, so that the error operators are orthogonal in the trace inner product. Since the representation is irreducible, the $E_g$ linearly span the set of all linear operators. It follows that the order of $G$ is the square of the dimension of the representation and therefore the $E_g$ form a nice error basis. Since $\sigma(H)$ is generated by $\sigma(Z)$ and the $E_g$, $H$ is isomorphic to an error group. $\square$

Theorem 2.1 implies that nice error bases can be obtained and classified by using group theoretic techniques. The problem of characterizing all groups with irreducible characters supported on the center remains open. The best known error basis, the two dimensional bit/sign flip error basis, satisfies that the index group $G$ is abelian. Sebastian Egner was the first to find an error basis with non-abelian index group. Here is a method for obtaining such error bases. This method generalizes the tensor product construction.

Let $H_1$ and $H_2$ be error groups on Hilbert spaces $\mathcal{H}_1$ and $\mathcal{H}_2$, respectively. Let $\phi$ be a homomorphism from $H_2$ to the normalizer of $H_1$ in the unitary group on $\mathcal{H}_1$. Note that $\phi$ is a representation of $H_2$ in $\mathcal{H}_1$. Assume that $\phi$ maps the center of $H_2$ to multiples of the identity. The map $\phi$ induces a homomorphism $\bar{\phi}$ into the automorphism group of $H_1$ by $\bar{\phi}(h)(x) = \phi(h) x \phi(h^{-1})$. This allows construction of an extension of $H_1$ by $H_2$, $H_1 \ltimes_{\bar\phi} H_2$. This extension is given by the set $H_1 \times H_2$ with the product $(h, g) * (h', g') = (h \bar{\phi}(g)(h'), gg')$. An irreducible representation of $H_1 \ltimes_{\bar\phi} H_2$ with the desired property is obtained by representing $(h, g)$ by $h\phi(g) \otimes g$. To see that this works, note that

$$\begin{aligned}(h\phi(g) \otimes g)(h'\phi(g') \otimes g' &= h\phi(g) h' \phi(g^{-1}) \phi(g) \phi(g') \otimes gg' \\ &= h\bar{\phi}(g)(h') \phi(gg') \otimes gg' \,.\end{aligned}$$

To show that the trace conditions for error groups are satisfied, write $\text{tr}(h\phi(g) \otimes g) = \text{tr}(h\phi(g)) \, \text{tr}(g)$. For this to be non-zero requires $\text{tr}(g) \neq 0$, so $g$ is a multiple of the identity. By assumption $\phi(g)$ is also proportional to the identity, so the trace is non-zero iff $\text{tr}(h) \neq 0$, which holds iff $h$ is a multiple of the identity. Thus the character of $H_1 \ltimes_{\bar\phi} H_2$ associated with the representation is supported on the center. The fact that it is irreducible follows by observing that the quotient over the center has sufficiently many elements to linearly generate any operator on $\mathcal{H}_1 \otimes \mathcal{H}_2$. That representatives of the quotient are linearly independent follows by orthogonality in the trace inner product.

Egner's error basis has the same quotient over the center as one constructed by this method. Let $N, S$ be the generators of the bit/sign flip error basis $\mathcal{E}_2$ as defined in [8]. The normalizer $\mathcal{N}$ of $\bar{\mathcal{E}}_2$ includes the Hadamard transform
$$H = \begin{pmatrix} 1 & 1 \\ 1 & -1 \end{pmatrix},$$
in addition to the elements of $\mathcal{E}_2$. A suitable homomorphism of $\mathcal{E}_2$ into $\mathcal{N}$ is obtained by setting $\phi(N) = \phi(S) = H$. The error basis obtained is generated by
$$A = NH \otimes N, \ B = S \otimes I, \ C = I \otimes NS.$$
These operators satisfy $A^4 = -I$, $B^2 = I$, $C^2 = -I$, $AC = -CA$, $BC = CB$, and $AB = -BA^{-1}$. It follows that the quotient over the center is isomorphic to $\mathbb{Z}_2 \times D_8$. Egner's error group is not isomorphic to the one constructed above.

## 3 Codes and Error Bases

Since error bases generate the algebra of linear operators on a Hilbert space, any physical effect on a state can be expressed as mixed sum of operators of an error basis [9]. Normally, one is interested in a tensor product space $\mathcal{H}^{\otimes n}$, and relevant effects tend to be local. Any error basis on $\mathcal{H}$ induces an error basis on $\mathcal{H}^{\otimes n}$ by the tensor product construction. The techniques of Calderbank et al. [2] exploit the structure of $\mathcal{E}_2^{\otimes n}$ to construct quantum codes which can correct suitably localized errors. The goal of this section is to generalize this technique to arbitrary error bases and cast it in representation theoretic terms.

Let $\mathcal{E}$ be a nice error basis on $\mathcal{H}$. Suppose that $\mathcal{E}$ generates a finite error group $\bar{\mathcal{E}}$. Let $D$ be a set of operators and $\mathcal{C}$ a $d$-dimensional subspace of $\mathcal{H}$. Let $\Pi_\mathcal{C}$ be the projection operator onto $\mathcal{C}$. Then $\mathcal{C}$ is a *D-detecting* code iff for every $g \in D$,
$$\Pi_\mathcal{C} g \Pi_\mathcal{C} \propto \Pi_\mathcal{C}.$$
Let $S$ be another set of operators. Then $\mathcal{C}$ is an *S-correcting* code iff it is an $S^\dagger S$-detecting code[1]. The idea of focusing on detectability rather than correctability and the relationship between the two notions is due to Bennett et al. [1]. The formalization in terms of projection operators is an immediate consequence of the conditions given in [7, 1].

---
[1] $S^\dagger S = \{A^\dagger B \mid A, B \in S\}$.

In the case where $\mathcal{E} = \mathcal{E}_0^{\otimes n}$, an interesting problem is to find codes which can correct all errors involving at most $e$ of the factors of the underlying space. In that case $S_e$ is taken to consist of tensor products of error operators in $\mathcal{E}_0$ with at most $e$ of the factors not the identity. An *e-error-correcting* code is one which is $S_e$-correcting.

In [2], error-correcting codes are obtained as "eigenspaces" of suitable abelian subgroups of $\bar{\mathcal{E}}$. Consider a normal subgroup $N$ of $\bar{\mathcal{E}}$. The first task is to determine a useful notion of an "eigenspace" for $N$. To do so, view $N$ as an abstract group with the induced representation on $\mathcal{H}$. Let $\chi$ be an irreducible character of $N$ which appears in this representation. All irreducible characters in the induced representation of $N$ are related to $\chi$ by conjugation (see below). The span $\mathcal{C}(\chi)$ of the irreducible invariant subspaces associated with $\chi$ is a candidate for a code. The projection operator for this subspace is given by

$$\bar{\chi} = \frac{d_\chi}{|N|} \sum_{g \in N} \bar{\chi}(g) g \,,$$

where $d_\chi$ is the dimension of $\chi$. Other candidates are obtained by writing $\bar{\chi} = \sum_{i=1}^{d_\chi} e_i$ where $e_i e_j = \delta_{i,j} e_i$ and $e_i^\dagger = e_i$. Each $e_i$ is a projection operator onto a subspace $\mathcal{C}(e_i)$ of the range of $\bar{\chi}$. The $e_i$ are primitive orthogonal idempotents of the character $\bar{\chi}$. Let $r$ be the multiplicity with which $\chi$ appears in the representation of $N$. Then the dimensions of $\mathcal{C}(\chi)$ and $\mathcal{C}(e_i)$ are $r d_\chi$ and $r$, respectively. $\mathcal{C}(e_i)$ can detect more errors than $\mathcal{C}(\chi)$.

An important problem is to determine a useful subset of the detectable errors. It is convenient to consider the normalizer $N_U(N)$ of $N$ in the unitary group $U = U(\mathcal{H})$ on $\mathcal{H}$. Thus $N \subseteq \bar{\mathcal{E}} \subseteq N_U(N)$ and $N_U(N)$ can be viewed as an abstract group with an irreducible unitary representation on $\mathcal{H}$.

We begin with a few observations from representation theory.

**Lemma 3.1.** *The elements of $N_U(N)$ permute the irreducible invariant subspaces of $N$. $\bar{\mathcal{E}}$ acts transitively on the irreducible invariant subspaces of $N$.*

This means that if $\mathcal{C}$ is an irreducible invariant subspace of $N$ and $g \in N_U(N)$, then $g\mathcal{C}$ is also an irreducible invariant subspace of $N$. If the character of $\mathcal{C}$ is $\chi$, then the character of $g\mathcal{C}$ is $g\chi g^{-1}$, defined by $(g\chi g^{-1})(h) = \chi(g^{-1}hg)$ (the expressions are chosen so as to make sense in the group algebra generated by $N$). The elements $g$ of $\mathrm{N}_U(N)$ for which $g\chi g^{-1} = \chi$ constitute the *inertia* subgroup $T(\chi)$ of $\chi$. Note that the inertia subgroup contains $N$. It generalizes the notion of $N^\perp$ introduced in [2].

The next two lemmas are useful consequences of the fact that $\bar{\mathcal{E}}$ acts transitively on the irreducible invariant subspaces of $N$.

**Lemma 3.2.** *All irreducible invariant subspaces of $N$ have the same dimension.*

**Lemma 3.3.** *Each distinct irreducible character of the induced representation of $N$ occurs with the same multiplicity.*

It follows that the number of distinct irreducible characters of $N$ induced by the representation on $\mathcal{H}$ is given by $r = |\bar{\mathcal{E}}/T(\chi)|$.

If $g$ is not in $T(\chi)$, then $g\mathcal{C}$ is an invariant subspace with a different irreducible character. Since the representation is unitary, it follows that $g\mathcal{C} \perp \mathcal{C}$. In fact, if the $\chi_i$ are the distinct irreducible characters of $N$, then $\bar{\chi}_i\bar{\chi}_j = \delta_{ij}\bar{\chi}_i$ and the $\bar{\chi}_i$ are Hermitian. Thus the $\bar{\chi}_i$ form a complete set of orthogonal projections. Thus $\bar{\chi}g\bar{\chi} = 0$ and $e_ige_i = 0$, and $g$ is detectable by both $\mathcal{C}(\chi)$ and $\mathcal{C}(e_i)$.

**Theorem 3.4.** *If $g$ is not in $T(\chi)$ or if $g$ is in $N$, then it is detectable by $\mathcal{C}(e_i)$. If $g$ is not in $T(\chi)$ or if $g$ is in the center of the irreducible representation of $\chi$, then it is detectable by $\mathcal{C}(\chi)$.*

*Proof.* The case of $g \notin T(\chi)$ has already been discussed. If $g$ is in the center of the irreducible representation, then in that representation it is a multiple of the identity, which implies $\bar{\chi}g\bar{\chi} \propto \bar{\chi}$. For $g \in N$, the orthogonal idempotents have the property that $e_ige_i \propto e_i$. This follows from the fact that in the Fourier transform of the group algebra the $e_i$ are diagonal matrices with only one non-zero entry on the diagonal. □

Theorem 3.4 allows constructing error-detecting and error-correcting codes from any normal subgroup of an error basis. Whether the greater generality helps with finding good codes remains to be determined.

The construction of [2] is based on using abelian normal subgroups $A$ of $\bar{\mathcal{E}}$ and making sure that the errors that need to be detected either are in $A$ or satisfy an anti-commutativity relationship with an element of $A$. This is justified as follows: Let $E$ be an error operator and assume that for some element $a \in A$, $aE = \omega Ea$, with $\omega \neq 1$ a scalar. Then $E^{-1}aE = \omega a$ so that $\chi(E^{-1}aE) = \omega\chi(a) \neq \chi(a)$. This implies that $E \notin T(\chi)$, so that $E$ is detectable. Note that for error groups which are commutative over the center, the elements of the inertia subgroup are exactly those which commute with every element of $A$. This simplifies the calculations substantially.

Consider the problem of finding a linear basis of the operators which are detectable by $\mathcal{C}(\chi)$. Let $r$ be the number of distinct irreducible characters induced in $N$. Then the dimension of $\mathcal{C}(\chi)$ is $n/r$. A simple counting argument based on representing linear operators in an extension of a basis of $\mathcal{C}(\chi)$ shows that the dimension of the space of detectable operators is $n^2 - (n^2/r^2 - 1)$. The number of independent elements of $\bar{\mathcal{E}} \setminus T(\chi)$ is $n^2 - n^2/r$. Let $\mathcal{D}$ be the set of operators which leave invariant the miminal invariant subspaces of $N$ and are multiples of the identity on $\mathcal{C}(\chi)$. The dimension of $\mathcal{D}$ is $n^2/r - n^2/r^2 + 1$, and these operators are orthogonal in the trace norm to the members of $\bar{\mathcal{E}} \setminus T(\chi)$. Furthermore, the operators of $\mathcal{D}$ are detectable. It follows that the detectable operators are linearly generated by $\mathcal{D} \cup \bar{\mathcal{E}} \setminus T(\chi)$.

## 4 Syndromes of Codes Based on Normal Subgroups

Let $\chi$ and $N$ be as in the previous section. Let $|\psi\rangle$ be an initial state in $\mathcal{C}(\chi)$. Let $|\psi'\rangle = \mathcal{A}|\psi\rangle$ be the state after an interaction with the environment. For error detection it suffices to measure whether $|\psi'\rangle$ is in $\mathcal{C}(\chi)$. If $\mathcal{A}$ is detectable and the outcome of the measurement is to project $|\psi'\rangle$ into $\mathcal{C}(\chi)$, then the resulting state is $|\psi\rangle$. All we can learn from the other measurement outcomes is that an error occurred.

To correct errors requires restoring $|\psi'\rangle$ according to a syndrome representation of the Hilbert space, $\mathcal{H} \simeq \mathcal{S} \otimes \mathcal{C}(\chi) + \mathcal{R}$. A suitable syndrome representation can be obtained from $\bar{\mathcal{E}}/T(\chi)$ as follows: Let $\{I = g_0, g_1, \ldots\}$ be a set of representatives in $\bar{\mathcal{E}}$ of $\bar{\mathcal{E}}/T(\chi)$[2]. Let $\mathcal{S}$ be the Hilbert space spanned by orthonormal states labeled by $|g_i\rangle$. Let $|j_L\rangle$ denote an orthonormal basis of $\mathcal{C}(\chi)$. An isomorphism $\sigma : \mathcal{S} \otimes \mathcal{C}(\chi) \to \mathcal{H}$ is established by defining

$$\sigma(|g_i\rangle|j_L\rangle) = g_i|j_L\rangle.$$

The choice of the $g_i$ implies that the subspaces $g_i\mathcal{C}(\chi)$ partition $\mathcal{H}$ into equal dimensional orthogonal subspaces (the invariant subspaces of $N$ associated with the distinct induced irreducible characters). To attempt to recover $|\psi\rangle$ from $|\psi'\rangle$, first measure the syndrome by projecting $|\psi'\rangle$ into one of the $g_i\mathcal{C}(\chi)$. If the outcome of the measurement is in $g_i\mathcal{C}(\chi)$, apply $g_i^{-1}$ to the state. Note that the $g_i\mathcal{C}(\chi)$ are independent of the choice of representatives $g_i \in g_iT(\chi)$. Only the restoration of the state to $\mathcal{C}(\chi)$ depends on this choice.

---

[2]For the present discussion, we consider only those elements of the inertia subgroup which are in $\bar{\mathcal{E}}$.

Given this procedure it is possible to determine which error operators in $\bar{\mathcal{E}}$ can be corrected. Let $E \in \bar{\mathcal{E}}$ and $g_i$ the representative of $ET(\chi)$. $E$ is correctable iff $g_i^{-1}E$ is in the center $C \subseteq N$ of the irreducible representation associated with $\chi$. It follows that the correctable errors in $\bar{\mathcal{E}}$ consist of the set $\bigcup_i g_i C$. Note that $C$ is a normal subgroup of $N$. Choosing $g_0 \neq I$ corresponds to using a biased recovery operation.

In most cases, the set of operators $S$ in $\bar{\mathcal{E}}$ which is intended to be correctable is given. Suppose that $S^\dagger S$ is detectable by $\mathcal{C}(\chi)$ and that a recovery procedure for correcting $S$ has to be designed. This is accomplished by choosing appropriate representatives $g_i$ of the cosets of $T(\chi)$. Suppose that $s_1, s_2 \in S$ belong to the same coset of $T(\chi)$ in $\bar{\mathcal{E}}$. Then $s_1^{-1}s_2 = s_1^\dagger s_2$ is a multiple of the identity on $\mathcal{C}(\chi)$ (by detectability and the fact that $s_1^{-1}s_2\mathcal{C}(\chi) = \mathcal{C}(\chi)$). Thus $s_1^{-1}s_2 \in C$, which means that $s_1$ and $s_2$ are in the same coset $gC$ of $C$. It suffices to choose the representative of $g'T(\chi)$ from among $S \cap g'T(\chi) \subseteq gC$ if $S \cap g'T(\chi) \neq \emptyset$. For other cosets, the representative can be chosen arbitrarily.

Similar arguments apply to $\mathcal{C}(e_1)$, if $e_1$ is a primitive orthogonal idempotent of $\bar{\chi}$. Given the decomposition $\bar{\chi} = \sum_i e_i$, let $e_{ij}$ be the elements in the group algebra of $N$ which satisfy $e_{ij}e_k e_{ij}^\dagger = \delta_{jk}e_i$ and $e_{ij}\bar{\chi}' = 0$ for every irreducible character $\chi'$ of $N$ different from $\chi$. The syndrome space is spanned by $|g_i, e_{j1}\rangle$ and we define

$$\sigma(|g_k, e_{j1}\rangle|l_L\rangle) = g_i e_{j1}|l_L\rangle,$$

where the $|l_L\rangle$ form an orthonormal basis of $\mathcal{C}(e_1)$. If the syndrome $g_i e_{j1}\mathcal{C}(e_1)$ is detected, the state is restored by applying (a unitary extension of) $e_{1j}g_i^{-1}$ to the result. In this case the correctable errors in $\bar{\mathcal{E}}$ consist of the set $\bigcup_i g_i N$.

## 5 The Group of Transversally Implementable Operations

Once a code on $\mathcal{H}^{\otimes n}$ has been constructed by the technique described in the previous sections, the inertia subgroup can be used to find operations which can be implemented transversally. This ensures that these operations can be implemented fault tolerantly. Assume that $\mathcal{C} = \mathcal{C}(\chi)$ for an irreducible character of the normal subgroup $N$ of $\mathcal{E}^{\otimes n}$. Let $\mathcal{H}$ be the Hilbert space acted on by $\mathcal{E}$ and $U(\mathcal{H})$ the unitary group on $\mathcal{H}$. The group of transversally implementable operations, $\mathcal{O}(\mathcal{C})$, consists of unitary operators of the form $U = U_1 \otimes \ldots \otimes U_n \in U(\mathcal{H})^{\otimes n}$ with $U\mathcal{C} = \mathcal{C}$. The question of which unitary operations on $\mathcal{C}$ occur in $\mathcal{O}(\mathcal{C})$ is of great interest when exploiting $\mathcal{C}$ for fault tolerant computation.

**Theorem 5.1.** *The inertia subgroup $T(\chi) \cap U(\mathcal{H})^{\otimes n}$ is contained in $\mathcal{O}(\mathcal{C})$.*

*Proof.* Since $g\chi g^{-1} = \chi$, $g$ commutes with $\bar{\chi}$, which implies that $g\mathcal{C} = \mathcal{C}$. □

If it is necessary to determine transversally encodable operations involving two encoded spaces, $\mathcal{C} \otimes \mathcal{C} \subset \mathcal{H}^{\otimes 2n}$, one can proceed as follows: First pair up the supporting systems as $(\mathcal{H} \otimes \mathcal{H})^{\otimes n}$, with the codes supported on the first and second member of each pair, respectively. $\mathcal{C} \otimes \mathcal{C}$ is given by $\mathcal{C}(\chi \otimes \chi)$. Note that $\chi \otimes \chi$ is an irreducible character of $N \times N$. Determine the intersection of the inertia subgroup of $\chi \otimes \chi$ with those elements of the normalizer of $N \times N$ which act on each pair independently. This provides a subset of two system transversally implementable operations.

# 6 Using Classical Linear Codes over $\mathrm{GF}(p^k)$

In [8] it was shown how to use general abelian error bases to exploit linear codes over $\mathbb{Z}_n$. The question of whether codes over non-prime fields could be used was left open. Let $\mathcal{H}$ be a $p$-dimensional Hilbert space with $p$ a prime. To exploit punctured codes over $\mathrm{GF}(p^k)$, consider $\mathrm{GF}(p^k)$ as a $k$-dimensional $\mathbb{Z}_p = \mathrm{GF}(p)$ vector space. Let $\omega$ be a primitive $p$-th root of unity and $b$ a $\mathbb{Z}_p$-linear form on $\mathrm{GF}(p^k)$. For $x, y \in \mathrm{GF}(p^k)$, denote the product in $\mathrm{GF}(p^k)$ by $x * y$. Thus $b(x * y)$ is a non-degenerate $\mathbb{Z}_p$-bilinear form. Define
$$C_y|x\rangle = |x+y\rangle, \quad D_y|x\rangle = \omega^{b(y*x)}|x\rangle.$$

These operators induce a nice error basis $\mathcal{E}_b$ on $\mathcal{H}^k$ with index group $\mathbb{Z}_p^{2k}$. It is therefore closely related to the standard tensor product error basis, but uses a different ordering of the classical basis. The generic element of $\mathcal{E}_b$ can be written as $E(x,y) = C_x D_y$. The elements can be multiplied by using the commutation relationship
$$C_x D_y = \omega^{-b(x*y)} D_y C_x.$$

Let $\mathcal{C}_0 \subseteq \mathcal{C}$ be codes over $\mathrm{GF}(p^k)^n$, where $\mathcal{C}_0$ has codimension one in $\mathcal{C}$. Thus $\mathcal{C} = \mathcal{C}_0 \uplus \ldots \uplus \mathcal{C}_{p-1}$, with $\mathcal{C}_1 = \mathcal{C}_0 + c$ and $\mathcal{C}_i = \mathcal{C}_0 + ic$. Let $|i_L\rangle = \sum_{x \in \mathcal{C}_0} |x + ic\rangle$. Then $|i_L\rangle$ is an invariant subspace of $\omega^j C_u D_v$ with eigenvalue (or character) one exactly if $u \in \mathcal{C}_0$ and $v \in \mathcal{C}^{\perp_b}$, the dual relative to $b$ of $\mathcal{C}$. Thus $\{|i_L\rangle\}_{i=0}^{p-1}$ can be obtained as the code associated with the 1-character of the (normal) subgroup of $\bar{\mathcal{E}}_b$ generated by those operators. The inertia subgroup in $\bar{\mathcal{E}}_b$ is generated by $C_u D_v$ with $u \in \mathcal{C}$ and $v \in \mathcal{C}_0^{\perp_b}$, and if the minimum weight (relative to $\mathrm{GF}(p^k)$) of both $\mathcal{C}$ and $\mathcal{C}_0^{\perp_b}$ is at least $2e + 1$, then the code can detect (correct) any errors in $\bar{\mathcal{E}}_b$ involving at most $2e$ ($e$) of the subsystems $\mathcal{H}^k$. Similar considerations apply if $\mathcal{C}_0$ has higher codimension in $\mathcal{C}$.

To relate this to the properties of the code over $\mathrm{GF}(p^k)^n$, it suffices to observe that for any code $\mathcal{D}$, $\mathcal{D}^{\perp_b} = \mathcal{D}^\perp$. We have

$$\mathcal{D}^{\perp_b} = \{x \mid \forall y \in \mathcal{D}, b(x * y) = 0\}.$$

Thus $\mathcal{D}^\perp \subseteq \mathcal{D}^{\perp_b}$. To see that these two sets are the same it suffices to compare dimensions over $\mathbb{Z}_p$. Let $d$ be the dimension of $\mathcal{D}$ over $\mathrm{GF}(p^k)$. Then the dimension of $\mathcal{D}$ over $\mathbb{Z}_p$ is $kd$. Because the bilinear form is non-degenerate, the dimension of $\mathcal{D}^{\perp_b}$ is $k(n-d)$. The dimension of $\mathcal{D}^\perp$ over $\mathrm{GF}(p^k)$ is $n-d$, and hence over $\mathbb{Z}_p$ it is also $k(n-d)$.

# 7 Conclusion

The construction of quantum codes based on abelian groups has been generalized to non-abelian error bases and cast in terms of the theory of group representations. The relevance of the inertia subgroups to the error-correcting properties of the codes, the recovery operation and the ability to implement operations transversally has been demonstrated. One problem not addressed here concerns the operations required to encode and decode the codes constructed in this fashion. How this can be done for $\mathcal{E}_2^{\otimes n}$ was briefly discussed in [2]. The other issue incompletely resolved by this report and its predecessor is how to perform fault tolerant recovery operations. For $\mathcal{E}_2^{\otimes n}$ a solution can be found in [4]. Finally, no interesting codes based on non-abelian error bases have been demonstrated in this work. The ultimate utility of such a general approach still remains to be determined.

# 8 Acknowledgements

Thanks to Thomas Beth, Markus Grassl and Wojciech Zurek for valuable discussions and to Sebastian Egner for providing the first example of a nice error basis with nonabelian index group.